%% file: main.tex
\documentclass[onefignum,onetabnum,onealgnum,onethmnum]{siamonline190516}
\usepackage{siunitx}
\usepackage{adjustbox}
\input{nd_preamble.tex}

\title{Fitting Splines to Axonal Arbors Quantifies Relationship between Branch Order and Geometry}

\author{Thomas L. Athey$^{1,\ast}$,
Jacopo Teneggi$^{1}$,
Joshua T. Vogelstein$^{1}$,
Daniel Tward$^{2,3}$,
Ulrich Mueller$^{4}$,
Michael I. Miller$^{1}$}

\begin{document}

\maketitle

\blfootnote{$^1$Kavli Neuroscience Discovery Institute, Department of Biomedical Engineering, Johns Hopkins University, Baltimore, MD, USA,
$^2$Department of Computational Medicine, University of California, Los Angeles , CA, USA,
$^3$Department of Neurology, University of California, Los Angeles , CA, USA,
$^4$Department of Neuroscience, Johns Hopkins University, Baltimore, MD, USA.
}
\blfootnote{
$ ^{*} $ corresponding author: tathey1@jhu.edu.
}

\thispagestyle{empty}

\noindent
\setcounter{tocdepth}{3}
\setcounter{secnumdepth}{3}

\vspace{-15pt}

\pagenumbering{arabic}
\setcounter{page}{1}

\input{content}

\section*{Conflict of Interest Statement}

MM own Anatomy Works with the arrangement being managed by Johns Hopkins
University in accordance with its conflict of interest policies. The
remaining authors declare that the research was conducted in the absence
of any commercial or financial relationships that could be construed as a
potential conflict of interest.

The funders had no role in study design, data collection and analysis,
decision to publish, or preparation of the manuscript.

\section*{Author Contributions}

MM and DT advised on the theoretical direction of the paper. UM advised on the data application experiments. JV advised on the presentation of the results. TA and JT designed the study, implemented the software, and managed the manuscript text/figures. All authors contributed to manuscript revision.

\section*{Funding} 

This work is supported by the National Institutes of Health grants RF1MH121539, \\
P41EB015909, R01NS086888, U19AG033655 and the National Science Foundation Grant 2014862.

\section*{Acknowledgments}
We thank the MouseLight team at HHMI Janelia for providing us with access to this data, and answering our questions about it.

\section*{Data Availability Statement}
The datasets analyzed for this study can be found in the \href{https://registry.opendata.aws/open-neurodata/}{Open Neurodata AWS account}: \url{s3://open-neurodata/brainlit/brain1_segments} and \url{s3://open-neurodata/brainlit/brain2_segments}. Our package \texttt{brainlit} provides examples of accessing this data. Specifically, instrutions on how to reproduce the figures found here can be found at \url{http://brainlit.neurodata.io/link_stubs/axon_geometry_readme_link.html}.

\vspace{5mm}
\bibliographystyle{unsrtnat}
\bibliography{citations}

\clearpage
\appendix

\input{appendix}

\end{document}

%% file: nd_preamble.tex
\usepackage{comment}
\usepackage{lipsum}
\usepackage[margin=1in, footskip=36pt]{geometry}

\usepackage{endnotes}

\usepackage{titlesec}
\usepackage{titletoc}
\usepackage{pgfplotstable}
\titleclass{\task}{straight}[\section]
\newcounter{task}
\renewcommand{\thetask}{\arabic{task}}
\titleformat{\task}[hang]
    {\normalfont\LARGE\bfseries}{Task \thetask:}{1em}{}

\titleformat*{\task}{\color{header1}\bfseries}

\titlecontents{task}
              [3.8em] 
              {}
              {\contentslabel{2.3em}}
              {\hspace*{-2.3em}}
              {\titlerule*[1pc]{.}\contentspage}

\titlespacing*{\section}{0ex}{1ex}{1ex}
\titlespacing*{\subsection}{0ex}{1ex}{1ex}
\titlespacing*{\subsubsection}{0ex}{1ex}{1ex}
\titlespacing*{\paragraph}{0ex}{1ex}{1ex}
\titlespacing*{\subparagraph}{0pt}{1ex}{1ex}
\titlespacing*{\task}{0em}{1ex}{1ex}

\newcommand\blfootnote[1]{%
  \begingroup
  \renewcommand\thefootnote{}\footnote{#1}%
  \addtocounter{footnote}{-1}%
  \endgroup
}

\usepackage[sort&compress,comma,square,numbers]{natbib}

\usepackage{amsfonts,amsmath,amssymb}
\usepackage{bbm}

\usepackage{multicol}
\usepackage{enumitem}
\setlist[enumerate]{wide, labelindent=1cm,  noitemsep}
\setlist[itemize]{noitemsep}
\setlist[description]{noitemsep}

\usepackage{hhline}

\usepackage{todonotes}
\RequirePackage{ulem}

\definecolor{tableheadcolor}{gray}{0.92}
\newcommand{\topline}{ %
        \arrayrulecolor{blue1}\specialrule{0.1em}{\abovetopsep}{0pt}%
        \arrayrulecolor{tableheadcolor}\specialrule{\belowrulesep}{0pt}{0pt}%
        \arrayrulecolor{blue1}}
\newcommand{\midtopline}{ %
        \arrayrulecolor{tableheadcolor}\specialrule{\aboverulesep}{0pt}{0pt}%
        \arrayrulecolor{blue1}\specialrule{\lightrulewidth}{0pt}{0pt}%
        \arrayrulecolor{white}\specialrule{\belowrulesep}{0pt}{0pt}%
        \arrayrulecolor{blue1}}
\newcommand{\bottomline}{ %
        \arrayrulecolor{white}\specialrule{\aboverulesep}{0pt}{0pt}%
        \arrayrulecolor{blue1} %
        \specialrule{\heavyrulewidth}{0pt}{\belowbottomsep}}%

\pgfplotstableset{normal/.style ={%
        header=true,
        string type,
        font=\small,
    	column type={p{.092\textwidth}},
        every head row/.style={
            before row={\topline\rowcolor{tableheadcolor}},
            after row={\midtopline}
        },
        every odd row/.style={
            before row={\rowcolor{blue1}}
        },
        every last row/.style={
            after row=\bottomline
        },
        col sep=&,
        row sep=\midrule
    }
}

\usepackage{multirow}

\usepackage{booktabs}
\usepackage{fancyhdr}
\usepackage{fullpage}

\RequirePackage{titlesec}
\RequirePackage[font={footnotesize},{singlelinecheck=false}]{caption}
\usepackage[T1]{fontenc}
\usepackage[scaled]{helvet}
\usepackage[scaled=1.10, med]{zlmtt}

\usepackage{algorithm}
\usepackage{algpseudocode}



%% file: content.tex
\begin{abstract} %
   Neuromorphology is crucial to identifying neuronal subtypes and understanding learning. It is also implicated in neurological disease.
   However, standard morphological analysis focuses on macroscopic features such as branching frequency and connectivity between regions, and often neglects the internal geometry of neurons. In this work, we treat neuron trace points as a sampling of differentiable curves and fit them with a set of branching B-splines. 
   We designed our representation with the Frenet-Serret formulas from differential geometry in mind. The Frenet-Serret formulas completely characterize smooth curves, and involve two parameters, curvature and torsion. Our representation makes it possible to compute these parameters from neuron traces in closed form. These parameters are defined continuously along the curve, in contrast to other parameters like tortuosity which depend on start and end points.
   We applied our method to a dataset of cortical projection neurons traced in two mouse brains, and found that the parameters are distributed differently between primary, collateral, and terminal axon branches, thus quantifying geometric differences between different components of an axonal arbor. The results agreed in both brains, further validating our representation. 
   The code used in this work can be readily applied to neuron traces in SWC format and is available in our open-source Python package \texttt{brainlit}: \url{http://brainlit.neurodata.io/}.
\end{abstract}
\vspace{-10pt}
\section{Introduction}

Not long after scientists like Ramon y Cajal started studying the nervous system with staining and microscopy, neuron morphology became a central topic in neuroscience \citep{parekh2013neuronal}. Morphology became the obvious way to organize neurons into categories such as pyramidal cells, Purkinje cells, and stellate cells. However, morphology is important not only for neuron subtyping, but in understanding learning and disease. For example, a now classic neuroscience experiment found altered morphology in geniculocortical axonal arbors in kittens whose eyes had been stitched shut upon birth \citep{antonini1993rapid}. Also, morphological changes have been associated with the gene underlying an inherited form of Parkinson's disease \citep{macleod2006familial}. Neuron morphology has been an important part of neuroscience for over a century, and remains so -- one of the \href{https://braininitiative.nih.gov/}{BRAIN} Initiative Cell Census Network's primary goals is to systematically characterize neuron morphology in the mammalian brain.

Currently, studying neuron morphology typically involves imaging one or more neurons, then tracing the cells and storing the traces in a digital format. Several recent initiatives have accumulated large datasets of neuron traces to facilitate morphology research. NeuroMorpho.Org, for example, hosts a total of over 140,000 neuron traces from a variety of animal species \citep{Ascoli9247}. These traces are typically stored as a list of vertices, each with some associated attributes including connections to other vertices.

Many scientists analyze neuron morphology by computing various summary features such as number of branch points, total length, and total encompassed volume. Neurolucida, a popular neuromorphology software, employs this technique. Another approach focuses on neuron topology, and uses metrics such as tree edit distance \citep{heumann2009tree}. However, both of these approaches neglect \textit{kinematic} geometry, or how the neuron travels through space. Tortuosity index is a summary feature that captures internal axon geometry, but this feature depends on the definition of start and end points, and cannot capture an axon's curvature at a single point.

In this work, we look at neuron traces through the lens of differential geometry. In particular, we establish a system of fitting interpolating splines to the neuron traces, and computing their curvature and torsion properties. To our knowledge, curvature and torsion have never been measured in neuron traces. We applied this method to cortical projection neuron traces from two mouse brains in the MouseLight dataset from HHMI Janelia \citep{winnubst2019reconstruction}. In both brains, we found different distributions of these properties between primary, collateral, and terminal axon segments. The code used in this work is available in our open-source Python package \texttt{brainlit}: \url{http://brainlit.neurodata.io/}. 

\section{Methods}

\subsection{Spline Fitting}

\begin{figure}[ht]
    \centering
    \includegraphics[width=\textwidth]{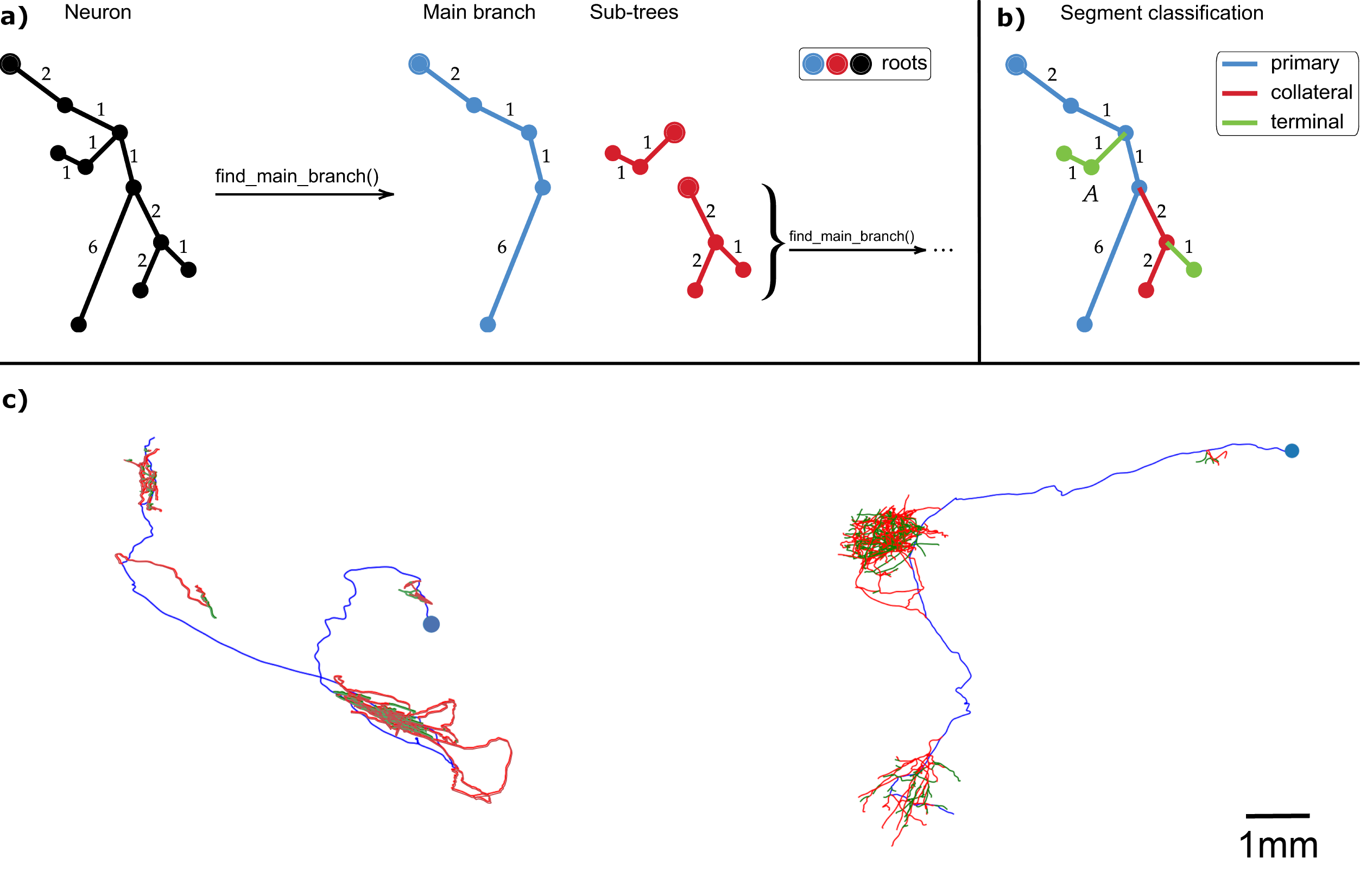}
    \caption{\textbf{a)} Illustration of how a neuron trace is split into different segments. This process identifies the longest root to leaf path (``Main branch''), and separates sub-trees from it. The sub-trees which still have branch points are processed in the same way until the neuron has been split into segments. By using path length to identify the Main branch, this splitting process is invariant to rigid transformations of the trace. \textbf{b)} Illustration of how axon segments are classified as primary, collateral, or terminal. The first segment is defined as primary, and segments that have no sub-trees are defined as terminal. All other segments are defined as collateral. \textbf{c)} Examples of splines that were fit to neuron traces. The splines pass through all trace points, and are thrice continuously differentiable for segments that contain at least five trace points. The blue points indicate the somas, and the spline colors indicate segment class (blue = primary, red = collateral, green = terminal). The neuron on the left is from brain 1, the one on the right is from brain 2.}
    \label{fig:spline}
\end{figure}

First, the neuron traces were split into segments by recursively identifying the longest root to leaf path (Figure~\ref{fig:spline}a). The first axon segment to be isolated in this way was defined to be the ``primary'' segment. Subsequent segments that branched were defined as ``collateral'' segments, and those that did not branch were defined to be ``terminal'' segments (Figure~\ref{fig:spline}b). This classification approximates the standard morphological definitions of primary, collateral and terminal axon branches.

Next, a B-spline was fit to each point sequence using \texttt{scipy}'s function \texttt{splprep} \citep{2020SciPy-NMeth}. \citet{kunoth2018splines} provide an in depth description of B-splines and their applications. Briefly, B-splines are linear combinations of piecewise polynomials, sometimes called basis functions. The basis functions are defined by a set of knots, which determine where the polynomial pieces meet, and degree, which determines the degree of the polynomial pieces. The $j$'th basis function for a set of knots $\mathbf{\xi}$ and degree $p$ is recursively defined by (Equation 1.1 in \citet{kunoth2018splines}):

\begin{align*}
    B_{j,p,\xi}&:=\frac{x-\xi_j}{\xi_{j+p}-\xi_j}B_{j,p-1,\mathbf{\xi}}(x)+\frac{\xi_{j+p+1}-x}{\xi_{j+p+1}-\xi_{j+1}}B_{j+1,p-1,\xi}(x) \\
    &\text{with} \\
    B_{i,0,\mathbf{\xi}}&:=\begin{cases}
    1, & \text{if } x\in [\xi_i,\xi_{i+1}), \\
    0,& \text{otherwise.}
    \end{cases}
\end{align*}

Splines are fit to data by solving a constrained optimization problem, where a smoothing term is minimized while keeping the residual error under a specified value \citep{dierckx1982algorithms}. Here, we constrain the splines to pass exactly through all points in the original trace, which corresponds to a smoothing condition of $s=0$ in \texttt{splprep}. For a sequence of $n > 5$ points, we fit a spline of degree $5$, which is the minimal degree that ensures that the splines are thrice continuously differentiable. Differentiability is important because it allows for estimation of curvature and torsion, explained in the next section.

Sequences of fewer than $5$ points, however, required lower degree splines to fully constrain the fitting procedure. For a sequence of $3 < n \leq 5$ points we used degree $3$, for a sequence of $n = 3$ points we used degree $2$, and for a sequence of $n=2$ points we used degree $1$. By selecting the degree in this way, we avoided splines of large even degree, such as fourth order splines, which are not recommended in our interpolation setting \citep{2020SciPy-NMeth}. Also, these degree choices are low enough to allow for a fully constrained fitting procedure, but high enough to make curvature/torsion nonvanishing when possible.

We recall that B-splines are not required to be parametrized by the arclength of the curve. Here, we set the $\mathbf{\xi} = \{0, \dots, L\}$, where $L$ is the cumulative length of the segments connecting the vertices of the trace, in $\mu m$. All other spline fitting options were set to the defaults in \texttt{splprep}. This spline fitting method can be applied to any set of points organized in a tree structure, such as a SWC file. Figure~\ref{fig:spline}c shows examples of splines that were fit to neuron traces.

\subsection{Frenet-Serret Parameters}
An important advantage of B-splines is that their derivatives can be computed in closed form. In fact, their derivatives are defined in terms of B-splines as shown below in Theorem 3 from \citet{kunoth2018splines}:

\textbf{Theorem} \textit{For a continuously differentiable b-spline $B_{j,p,\mathbf{\xi}}(\cdot)$ defined by index $j$, degree $p\geq 1$, and knot sequence $\mathbf{\xi}$, we have:}

\begin{align*}
    \frac{d}{ds}B_{j,p,\mathbf{\xi}}(s)=p\left( \frac{B_{j,p-1,\mathbf{\xi}}(s)}{\xi_{j+p}-\xi_j}-\frac{B_{j+1,p-1,\mathbf{\xi}}(s)}{\xi_{j+p+1}-\xi_{j+1}}\right)
\end{align*}

\textit{where we assume by convention that fractions with zero denominator have value zero.}

Curvature and torsion can be easily computed because of this property. For a thrice differentiable curve $x(s) \in \mathbb{R}^3$ that is parameterized by arclength (i.e. $||\dot x (s)||=1 \; \forall s$), one can compute the curvature ($\kappa$) and torsion ($\tau$) with the following formulas:

\begin{align*}
    \kappa(s)&=||\dot x(s) \times \ddot x(s)|| \\
    \tau (s) &= \frac{\langle (\dot x(s) \times \ddot x(s)), \dddot x(s) \rangle}{||\dot x(s) \times \ddot x(s)||^2}
\end{align*}

\noindent defined with the standard Euclidean norm $||\cdot||$, inner product $\langle \cdot ,\cdot \rangle$, and cross product $\times$. When curvature vanishes, we define torsion to be zero as well, since the torsion equation becomes undefined. The units of curvature and torsion are both inverse length. In this work, neuron traces have units of microns, so curvature and torsion both have units of $(\mu m)^{-1}$.

Curvature measures how much a curve deviates from being straight, and torsion measures how much a curve deviates from being planar. Together, these quantities parametrize the Frenet-Serret formulas of differential geometry. These formulas completely characterize continuously differentiable curves in three-dimensional Euclidean space, up to rigid motion \citep{grenander2007pattern}. Curvature takes non-negative values, but torsion can be positive or negative where the sign denotes the direction of the torsion in the right-handed coordinate system. In this work, we are not interested in the direction of the torsion, so we focused on the torsion magnitude (absolute value).

\subsection{Data}

We applied our methods to a collection of cortical projection neuron axon traces from two mouse brains in the HHMI Janelia MouseLight dataset. The precision of the reconstructions is limited by the resolution of the original two-photon block-face images, which was $0.3 \mu m \times 0.3 \mu m \times 1 \mu m$ \citep{winnubst2019reconstruction}. Each reconstruction was reviewed by two independent annotators to ensure that the reconstructions were accurate. There were 180 traces from brain 1 and 50 traces from brain 2. 

After fitting splines to these traces, curvature and torsion magnitude were sampled every $1 \mu m$ along the axon segments. We chose this sampling frequency because $1 \mu m$ is the z-resolution of the original brain images, so it is not likely that any higher sampling frequency would add any meaningful information. We studied curvature and torsion magnitude in two ways, described below in Sections~\ref{sec:auto_method} and~\ref{sec:seg_method}. 

\subsection{Computing Autocorrelation of Curvature and Torsion}
\label{sec:auto_method}

Our first goal was to identify the length scale at which straight axon segments remain straight and curved axon segments remain curved, so we studied the autocorrelation of curvature and torsion magnitude along the axon segments. For each axon segment, the autocorrelation functions of curvature and torsion were computed along the length of the segment, yielding a collection of autocorrelation functions for each brain. Then, we evaluated whether autocorrelation at a particular lag was significantly higher than $0.3$ using a one-sided t-test with a significance threshold of $\alpha = 0.05$. We identified $0.3$ as our effect size because correlations higher than $0.3$ are generally regarded as ``moderate'' correlations.

It is worth noting that, by the nature of the spline fitting procedure in \citet{2020SciPy-NMeth}, ``lag'' in our autocorrelation functions refers to straight line distances between the trace points, not by the arclength of the resulting curves.

\subsection{Comparing Axon Segment Classes}
\label{sec:seg_method}

Our second goal in the analysis was to compare curvature/torsion between segment classes. First, we estimated each segment's average curvature/torsion magnitude by taking the mean from all points that were sampled on that segment.

In order to compare different segment classes, we developed a paired sample method for testing for differences in average curvature/torsion. Different neurons represented different samples, and the average curvature/torsion of two segment classes (primary vs. collateral, collateral vs. terminal, primary vs. terminal) represented the paired measurements.

Define the random variable $X$ as the average curvature/torsion of one segment class and $Y$ as the average curvature/torsion of another segment class. Further, say $X$ and $Y$ are both real valued. Our null and alternative hypotheses are as follows:

\begin{align*}
    \mathcal{H}_0   &: \Pr[X>Y] = 0.5\\
    \mathcal{H}_1   &: \Pr[X>Y] \neq 0.5
\end{align*}

We tested these hypotheses using the sign test \citep{neuhauser2011nonparametric}. The test statistic is the number of times that the data point from one sample is greater than its pair from the other sample. A key advantage of the sign test is that it does not require parametric distribution assumptions, such as normality of the data. Also, its null distribution can be computed exactly via the binomial distribution. The two different parameters (curvature and torsion), and the three different segment class pairs constitute six total tests, so we applied the Bonferroni correction to $\alpha=0.05$ to obtain the significance threshold $0.0083$, which controls the family-wise error rate to $0.05$. We conducted one-sided sign tests in all cases.

\section{Results}
\subsection{Autocorrelation of Curvature and Torsion}
\label{sec:autocorr}

\begin{figure}[ht] 
    \centering
    \includegraphics[width=\textwidth]{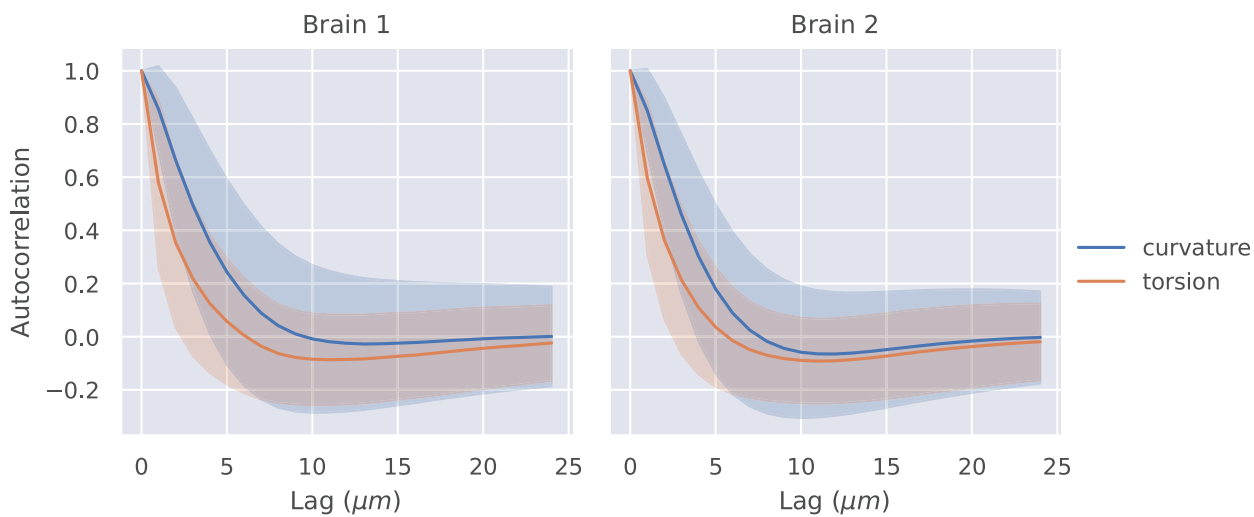}
    \caption{Autocorrelation of curvature and torsion magnitude averaged across all axon segments with $\pm 1\sigma$ confidence intervals. Curvature and torsion were sampled at every $1 \mu m$ along the axon segments. One sided t-tests indicated that curvature had statistically significant autocorrelation values above $0.3$ at lags of $1\mbox{-}4 \mu m$ in brain 1 and $1\mbox{-}3 \mu m$ in brain 2. Torsion had statistically significant autocorrelation values above $0.3$ at lags of $1$ and $2 \mu m$ in both brain 1 and 2.}
    \label{fig:autocor}
\end{figure}

The autocorrelation functions for all segments of a brain were averaged, and they are shown in Figure~\ref{fig:autocor}. Also shown is a shaded region that represents one standard deviation of these autocorrelation functions. The t-tests described in Section~\ref{sec:auto_method} were significant up to a lag of $4 \mu m$ for curvature in brain 1, $3 \mu m$ for curvature in brain 2, $2 \mu m$ for torsion in brain 1, and $2 \mu m$ for torsion in brain 2.

\subsection{Axon Segment Class Differences}
\label{sec:segments}

\begin{figure}[ht]
    \centering
    \includegraphics[width=\textwidth]{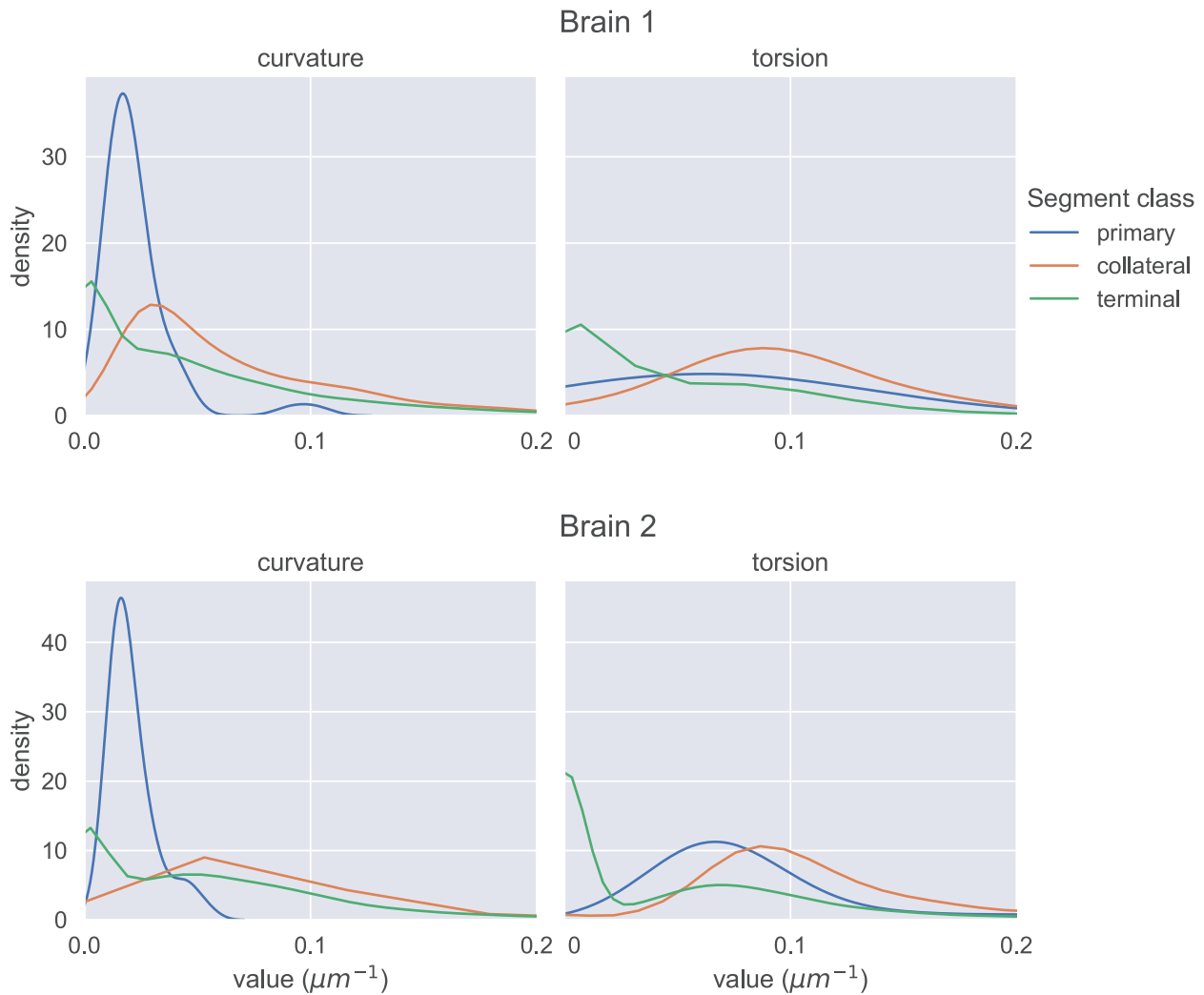}
    \caption{The distributions of average curvature and average torsion differed between the different segment classes as shown in these kernel density estimates, using a Gaussian kernel with a standard deviation of $2$. Segment averages were computed by sampling the curves at a uniform spacing of $1 \mu m$. One-sided sign tests, testing for differences in average curvature and torsion, were conducted while controlling the family-wise error rate to 0.05. The tests were significant in all cases and the directionality of the tests agreed in both brains.}
    \label{fig:ct_dists}
\end{figure}

The distributions of mean curvature and torsion are shown in Figure~\ref{fig:ct_dists}. Our statistical testing procedure, described in Section~\ref{sec:seg_method}, rejected the null hypothesis in all cases, with all p-values less than $5 \times 10^{-7}$. The directions of the one-sided tests were identical in both brains with:

\begin{align*}
    \text{Curvature: } \text{Collateral} > \text{Terminal} > \text{Primary} \\ 
    \text{Torsion: } \text{Collateral} > \text{Primary} > \text{Terminal} 
\end{align*}

\begin{figure}[ht]
    \centering
    \includegraphics[width=\textwidth]{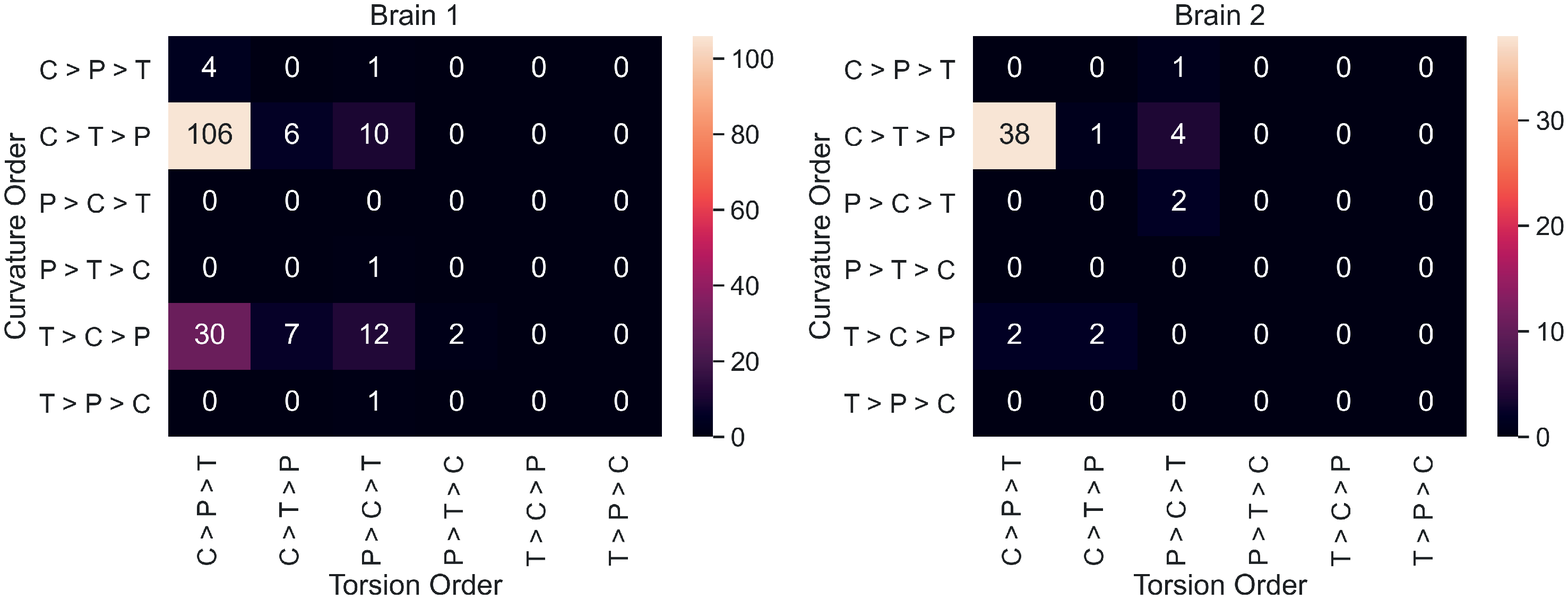}
    \caption{For each neuron, average curvature and torsion was computed for all three segment classes (P=primary, C=collateral, T=terminal) and compared between classes.  These heatmaps show the neuron counts for all 36 possible orderings of curvature/torsion. The most common ordering was collateral $>$ terminal $>$ primary for curvature and collateral $>$ primary $>$ terminal for torsion.}
    \label{fig:heatmaps}
\end{figure}

Neuron counts for all 36 possible curvature/torsion orderings across classes are shown in Figure~\ref{fig:heatmaps}. The most common ordering of curvature/torsion is exactly the same as the results of the sign test (90/180 neurons followed this ordering in brain 1, 30/50 in brain 2).

In the Appendix, we plot the curvature/torsion versus segment length. There appear to be modest correlations between segment length and curvature/torsion values in log-log plots.

\section{Discussion}

Our work proposes a model of neuron morphology using continuously differentiable B-splines. From these curves, it is possible to measure kinematic properties of neuronal processes, including curvature and torsion. These techniques are freely available in our open source Python package \texttt{brainlit}: \url{http://brainlit.neurodata.io/}, and more information about how to reproduce the specific results here can be found in the data availability statement.

In most contemporary neuromorphological analysis, neuron traces are regarded as piecewise linear structures, which precludes any analysis of higher order derivatives. Our spline representation makes it possible to estimate higher order derivatives and study parameters like curvature and torsion of neuron branches. In the popular piecewise linear representation, curvature and torsion would be zero along the line segments, and undefined where the line segments meet. We simulated a piecewise linear representation by modifying our spline fitting procedure to only produce splines of degree one. Indeed, with this less sophisticated representation, curvature and torsion vanished everywhere, making them not meaningful. 

Tortuosity index captures similar information to our curvature/torsion measurements and is popular in neuromorphological analysis \citep{STEPANYANTS2004251}. However, tortuosity requires the user to define start and end points whereas our method does not. Further, the piecewise linear representation of neuron traces limits the sampling frequency of tortuosity. Since tortuosity of a straight line is identically 1, placing the start and endpoints on the same linear segment will always produce a tortuosity value of 1. Our method, on the other hand, can produce more meaningful instantaneous curvature/torsion values.

Our methods for fitting splines and measuring curvature and torsion can be applied in neuromorphological analysis in a variety of ways, but we highlight two applications here, on a dataset of 230 projection neuron traces from two different mouse brains. We found that the autocorrelation functions of both curvature and torsion showed statistically significant correlations above $0.3$ within lags of approximately $2$ microns (specific lag values given in Section \ref{sec:autocorr}). Next, we defined segments as either ``primary,'' ``collateral,'' or ``terminal,'' and found significant differences in the distributions of curvature and torsion between these classes.

The statistical analysis approach described in Section~\ref{sec:seg_method} satisfies two desirable properties. First, by averaging measurements across segment classes, and pairing the data, we did not have to assume independence between segments of the same neuron. Assuming independence seemed inappropriate because, for example, segments that are connected to each other may have correlated geometry. Second, it avoided any parametric assumptions of the data, such as assuming normality of curvature/torsion measurements. A normality assumption seemed inappropriate for several reasons, including the fact that curvature is nonnegative, and that curvature/torsion was identically 0 for short segments with only 2 trace points.

Figure \ref{fig:heatmaps} shows that most individual neurons agree with the overall trend that collateral segments have the highest curvature and torsion. This suggests that the finding here is a consistent phenomenon among projection neurons in mice. In order to explore curvature/torsion distributions one level deeper, we looked into the relationship between curvature/torsion and segment length (see Supplement). In all segment classes, longer segments tend to have less curvature. The relationship between segment length and torsion is weaker, but there does appear to be a positive correlation.

Together, these findings suggest that the geometry of primary axon branches is different than that of higher order branches, such as the segments in terminal arborizations. In particular, higher order branches (collaterals and terminals) had higher curvature than primary branches. Collateral branches also had the highest torsion, but primary branches had higher torsion than terminal segments.

The primary limitation of our work is that our process of splitting a neuron trace into segments is based on segment length, and may not exactly partition the neuron into primary, collateral, and terminal branches according to their standard morphological definitions. Typically, collaterals are defined as branches that split off their parent branch at sharp angles, and arborize in a different location from other branches \citep{rockland2013collateral}. Future work could include changing our definitions of these classes to more closely reflect morphological definitions from scientific literature. Also, extending these experiments to neuron trace repositories such as NeuroMorpho.Org would help verify if our findings generalize.

Previous research has already indicated differences in axon geometry across neuronal cell types. For example, \citet{STEPANYANTS2004251} found higher tortuosity in the axons of GABAergic interneurons versus those of pyramidal cells. Similarly, \citet{portera2005diverse} found Cajal-Retzius cells to be significantly more tortuous than Thalamocortical (TC) cells, which is a type of projection neuron. \citet{portera2005diverse} also offers evidence that, while the primary axon in TC cells travel via a growth cone, most branching occurs via an interstitial, growth cone independent process. Our work elaborates on this distinction, suggesting that higher order axon branches have different geometry as well. While earlier research studied the differences of axonal geometry between neurons, we studied the variation of axonal geometry within neurons.

It is also worth noting that this is not the first work to model neuron traces as continuous curves in $\mathbb{R}^3$. For example, \citet{duncan2018statistical} construct a sophisticated and elegant representation of neurons that offers several useful properties. First, their representation is invariant to rigid motion and reparameterization. Second, their representation offers a vector space with a shape metric amenable to clustering and classification. However, their representation is limited to neuron topologies consisting of a main branch and only first order collaterals. Our B-splines approach does not immediately yield vector space properties, but can be applied to neurons with higher order branching, and allows for closed form computation of curvature and torsion. In short, the representation in \citet{duncan2018statistical} is designed for analysis between neurons, and our representation is designed for analysis within neurons. In the future, we are interested in bringing the advantages of their work to the open source software community, and combining it with the advantages of ours.

It is well known that axons are pruned and modified over time \citep{portera2005diverse}. It is possible that this process contributes to the different geometry of proximal versus distal axonal segments. Indeed, \citet{portera2005diverse} mentions the growth of short twisted branches towards the end of axon development. Future animal experiments could follow-up on this idea, and similar experiments to this one could be applied to other neuron types and other species to see if this is a widespread phenomenon in neuron morphology.

%% file: appendix.tex
\section{Supplementary Figures}

\begin{figure}[ht]
    \centering
    \includegraphics[width=0.85\textwidth]{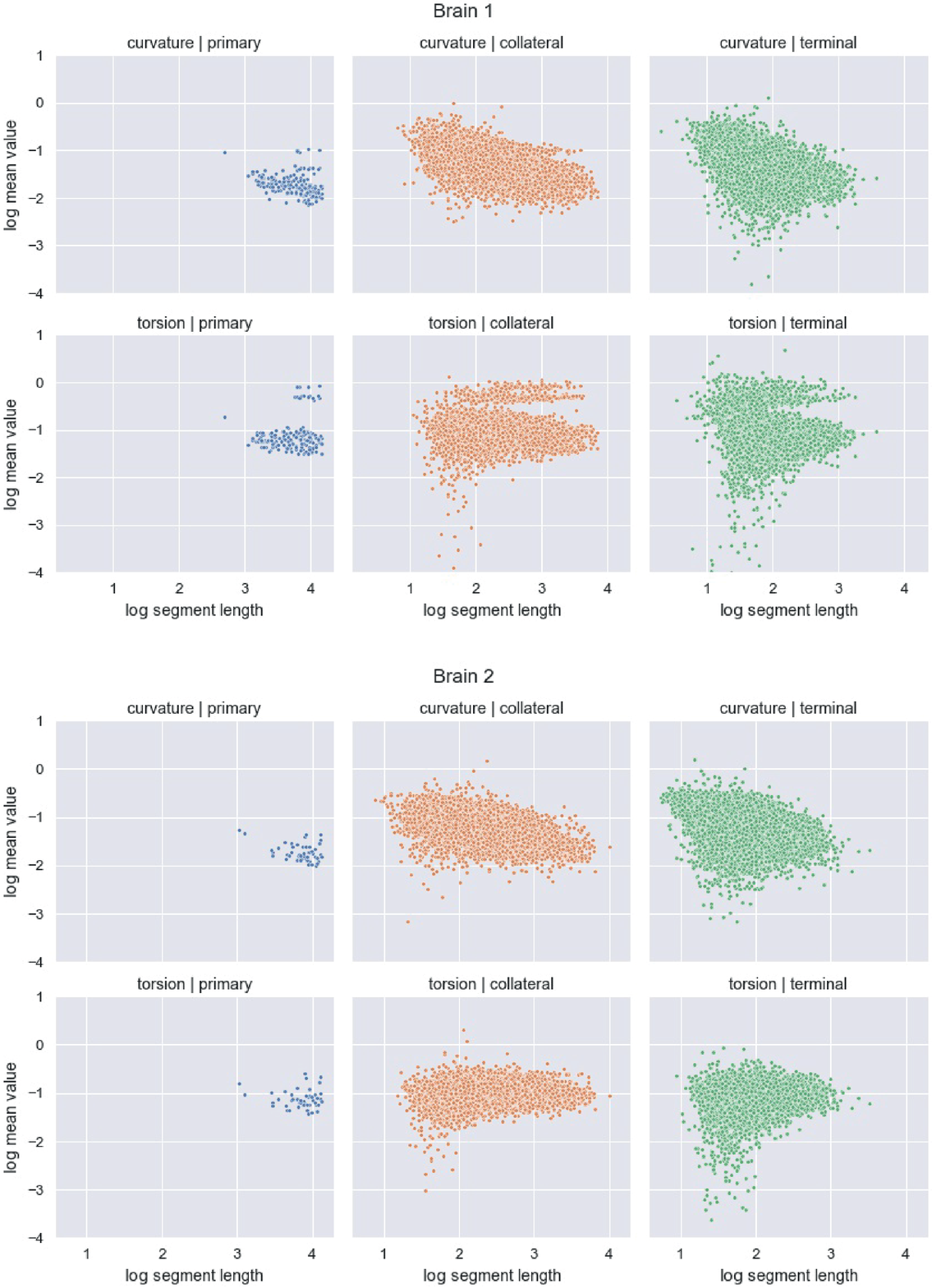}
    \caption{The above plots show the relationship between segment length, and mean curvature or torsion in each segment class and brain. Each data point represents a single axon segment, and average curvature and torsion was computed by sampling the segments at a uniform spacing of $14 \mu m$. We removed segments with zero average curvature/torsion in order to plot the data on a log scale. In this data, there appear to be weak negative correlations between segment length and curvature, and a weak positive correlations between segment length torsion.}
    \label{fig:mean_scatter}
\end{figure}

\begin{figure}[ht]
    \centering
    \includegraphics[width=\textwidth]{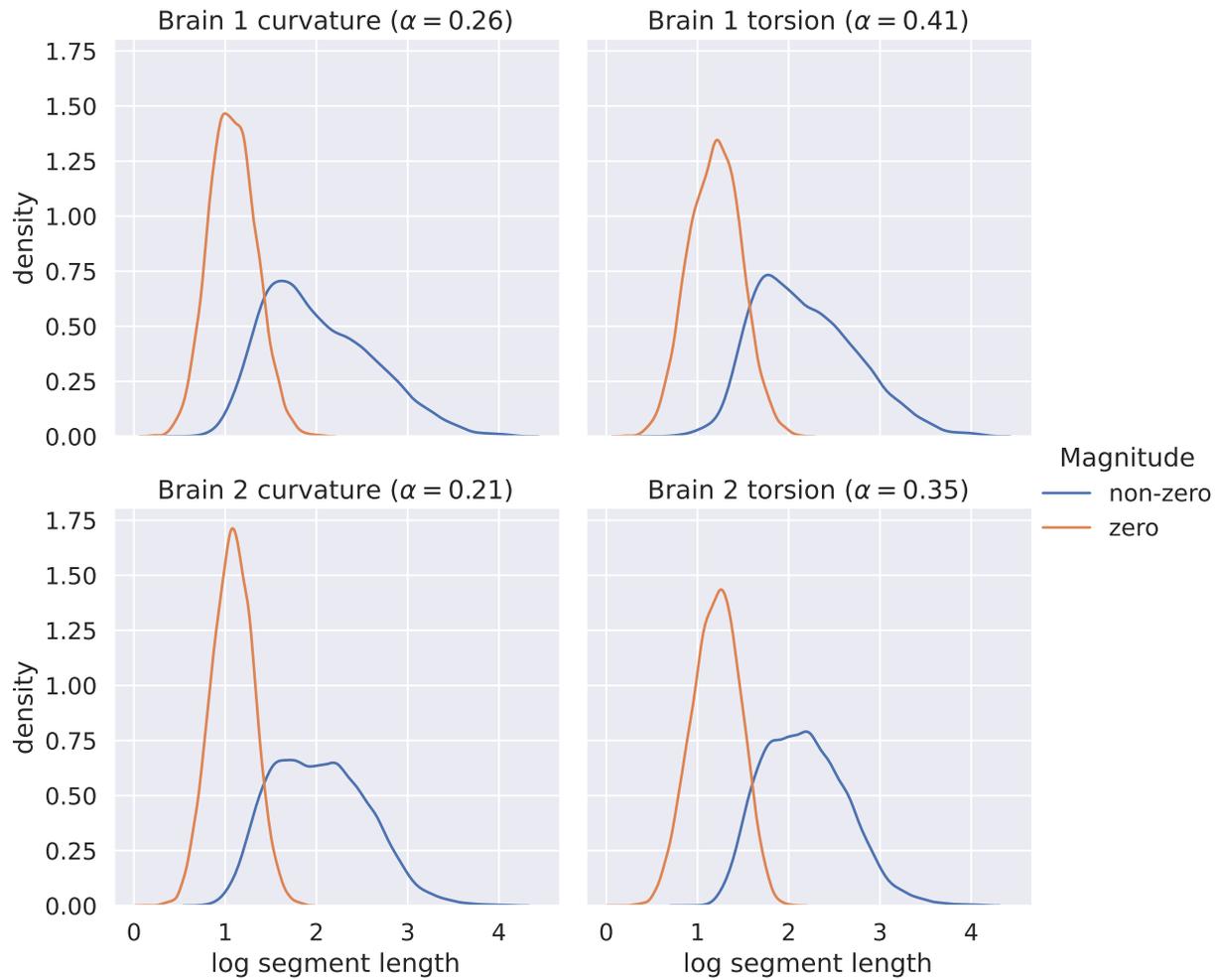}
    \caption{Segments with zero curvature or torsion were typically the shortest of all the segments. In almost all cases, segments had zero curvature/torsion because the segment was only composed of two or three trace points. $\alpha$ indicates the fraction of segments that had zero curvature/torsion in each brain.}
    \label{fig:dist}
\end{figure}

%% file: main.bbl
\begin{thebibliography}{15}
\providecommand{\natexlab}[1]{#1}
\providecommand{\url}[1]{\texttt{#1}}
\expandafter\ifx\csname urlstyle\endcsname\relax
  \providecommand{\doi}[1]{doi: #1}\else
  \providecommand{\doi}{doi: \begingroup \urlstyle{rm}\Url}\fi

\bibitem[Parekh and Ascoli(2013)]{parekh2013neuronal}
Ruchi Parekh and Giorgio~A Ascoli.
\newblock Neuronal morphology goes digital: a research hub for cellular and
  system neuroscience.
\newblock \emph{Neuron}, 77\penalty0 (6):\penalty0 1017--1038, 2013.

\bibitem[Antonini and Stryker(1993)]{antonini1993rapid}
Antonella Antonini and Michael~P Stryker.
\newblock Rapid remodeling of axonal arbors in the visual cortex.
\newblock \emph{Science}, 260\penalty0 (5115):\penalty0 1819--1821, 1993.

\bibitem[MacLeod et~al.(2006)MacLeod, Dowman, Hammond, Leete, Inoue, and
  Abeliovich]{macleod2006familial}
David MacLeod, Julia Dowman, Rachel Hammond, Thomas Leete, Keiichi Inoue, and
  Asa Abeliovich.
\newblock The familial parkinsonism gene lrrk2 regulates neurite process
  morphology.
\newblock \emph{Neuron}, 52\penalty0 (4):\penalty0 587--593, 2006.

\bibitem[Ascoli et~al.(2007)Ascoli, Donohue, and Halavi]{Ascoli9247}
Giorgio~A. Ascoli, Duncan~E. Donohue, and Maryam Halavi.
\newblock Neuromorpho.org: A central resource for neuronal morphologies.
\newblock \emph{Journal of Neuroscience}, 27\penalty0 (35):\penalty0
  9247--9251, 2007.
\newblock ISSN 0270-6474.
\newblock \doi{10.1523/JNEUROSCI.2055-07.2007}.
\newblock URL \url{https://www.jneurosci.org/content/27/35/9247}.

\bibitem[Heumann and Wittum(2009)]{heumann2009tree}
Holger Heumann and Gabriel Wittum.
\newblock The tree-edit-distance, a measure for quantifying neuronal
  morphology.
\newblock \emph{Neuroinformatics}, 7\penalty0 (3):\penalty0 179--190, 2009.

\bibitem[Winnubst et~al.(2019)Winnubst, Bas, Ferreira, Wu, Economo, Edson,
  Arthur, Bruns, Rokicki, Schauder, et~al.]{winnubst2019reconstruction}
Johan Winnubst, Erhan Bas, Tiago~A Ferreira, Zhuhao Wu, Michael~N Economo,
  Patrick Edson, Ben~J Arthur, Christopher Bruns, Konrad Rokicki, David
  Schauder, et~al.
\newblock Reconstruction of 1,000 projection neurons reveals new cell types and
  organization of long-range connectivity in the mouse brain.
\newblock \emph{Cell}, 179\penalty0 (1):\penalty0 268--281, 2019.

\bibitem[Virtanen et~al.(2020)Virtanen, Gommers, Oliphant, Haberland, Reddy,
  Cournapeau, Burovski, Peterson, Weckesser, Bright, {van der Walt}, Brett,
  Wilson, Millman, Mayorov, Nelson, Jones, Kern, Larson, Carey, Polat, Feng,
  Moore, {VanderPlas}, Laxalde, Perktold, Cimrman, Henriksen, Quintero, Harris,
  Archibald, Ribeiro, Pedregosa, {van Mulbregt}, and {SciPy 1.0
  Contributors}]{2020SciPy-NMeth}
Pauli Virtanen, Ralf Gommers, Travis~E. Oliphant, Matt Haberland, Tyler Reddy,
  David Cournapeau, Evgeni Burovski, Pearu Peterson, Warren Weckesser, Jonathan
  Bright, St{\'e}fan~J. {van der Walt}, Matthew Brett, Joshua Wilson, K.~Jarrod
  Millman, Nikolay Mayorov, Andrew R.~J. Nelson, Eric Jones, Robert Kern, Eric
  Larson, C~J Carey, {\.I}lhan Polat, Yu~Feng, Eric~W. Moore, Jake
  {VanderPlas}, Denis Laxalde, Josef Perktold, Robert Cimrman, Ian Henriksen,
  E.~A. Quintero, Charles~R. Harris, Anne~M. Archibald, Ant{\^o}nio~H. Ribeiro,
  Fabian Pedregosa, Paul {van Mulbregt}, and {SciPy 1.0 Contributors}.
\newblock {{SciPy} 1.0: Fundamental Algorithms for Scientific Computing in
  Python}.
\newblock \emph{Nature Methods}, 17:\penalty0 261--272, 2020.
\newblock \doi{10.1038/s41592-019-0686-2}.

\bibitem[Kunoth et~al.(2018)Kunoth, Lyche, Sangalli, and
  Serra-Capizzano]{kunoth2018splines}
Angela Kunoth, Tom Lyche, Giancarlo Sangalli, and Stefano Serra-Capizzano.
\newblock \emph{Splines and PDEs: From approximation theory to numerical linear
  algebra}.
\newblock Springer, 2018.

\bibitem[Dierckx(1982)]{dierckx1982algorithms}
Paul Dierckx.
\newblock Algorithms for smoothing data with periodic and parametric splines.
\newblock \emph{Computer Graphics and Image Processing}, 20\penalty0
  (2):\penalty0 171--184, 1982.

\bibitem[Grenander et~al.(2007)Grenander, Miller, Miller,
  et~al.]{grenander2007pattern}
Ulf Grenander, Michael~I Miller, Michael Miller, et~al.
\newblock \emph{Pattern theory: from representation to inference}.
\newblock Oxford university press, 2007.

\bibitem[Neuhauser(2011)]{neuhauser2011nonparametric}
Markus Neuhauser.
\newblock \emph{Nonparametric statistical tests: A computational approach}.
\newblock CRC Press, 2011.

\bibitem[Stepanyants et~al.(2004)Stepanyants, Tamás, and
  Chklovskii]{STEPANYANTS2004251}
Armen Stepanyants, Gábor Tamás, and Dmitri~B Chklovskii.
\newblock Class-specific features of neuronal wiring.
\newblock \emph{Neuron}, 43\penalty0 (2):\penalty0 251--259, 2004.
\newblock ISSN 0896-6273.
\newblock \doi{https://doi.org/10.1016/j.neuron.2004.06.013}.
\newblock URL
  \url{https://www.sciencedirect.com/science/article/pii/S0896627304003629}.

\bibitem[Rockland(2013)]{rockland2013collateral}
Kathleen~S Rockland.
\newblock Collateral branching of long-distance cortical projections in monkey.
\newblock \emph{Journal of Comparative Neurology}, 521\penalty0 (18):\penalty0
  4112--4123, 2013.

\bibitem[Portera-Cailliau et~al.(2005)Portera-Cailliau, Weimer, De~Paola,
  Caroni, and Svoboda]{portera2005diverse}
Carlos Portera-Cailliau, Robby~M Weimer, Vincenzo De~Paola, Pico Caroni, and
  Karel Svoboda.
\newblock Diverse modes of axon elaboration in the developing neocortex.
\newblock \emph{PLoS Biol}, 3\penalty0 (8):\penalty0 e272, 2005.

\bibitem[Duncan et~al.(2018)Duncan, Klassen, Srivastava,
  et~al.]{duncan2018statistical}
Adam Duncan, Eric Klassen, Anuj Srivastava, et~al.
\newblock Statistical shape analysis of simplified neuronal trees.
\newblock \emph{Annals of Applied Statistics}, 12\penalty0 (3):\penalty0
  1385--1421, 2018.

\end{thebibliography}
